\documentclass[9pt,twocolumn,twoside]{opticajnl}
\journal{opticajournal} 

\setboolean{shortarticle}{true}

\usepackage{lineno}
\usepackage{upgreek}
\usepackage{stfloats}
\linenumbers 

\title{Observation of tuning properties in a doubly resonant backward optical parametric oscillator}

\author[1,2]{Ming-Yuan Gao}
\author[1,2]{Yue-Wei Song}
\author[1,2,3]{Zhi-Cheng Guo}
\author[1,2]{Yin-Hai Li}
\author[1,2,3,*]{Zhi-Yuan Zhou}
\author[1,2,3,**]{Bao-Sen Shi}

\affil[1]{CAS Key Laboratory of Quantum Information, University of Science and Technology of China, Hefei, Anhui 230026, China}
\affil[2]{CAS Center for Excellence in Quantum Information and Quantum Physics, University of Science and Technology of China, Hefei 230026, China}
\affil[3]{Hefei National Laboratory, University of Science and Technology of China, Hefei 230088, China}

\affil[*]{Corresponding author: zyzhouphy@ustc.edu.cn} 
\affil[**]{Corresponding author: drshi@ustc.edu.cn}

\begin{abstract}
 Doubly resonant optical parametric oscillators (OPOs) under continuous wave (CW) pumping are particularly notable for their low threshold and narrow linewidth. Backward OPOs (BOPOs) realized through backward quasi-phase matching which exhibit unique tuning properties compared with conventional forward OPOs have been demonstrated under pulse pumping. In this work, a doubly resonant BOPO was implemented in a semi-monolithic cavity under CW pumping, and its tuning properties were characterized. By tuning the pump wavelength, the forward and backward waves exhibited tuning ranges of 56.85 nm and 0.89 nm, respectively. Adjusting the crystal temperature resulted in tuning ranges of 59.7 GHz and 59.4 GHz for the forward and backward waves, respectively. This research establishes the BOPO as a promising candidate for applications in the field of CW OPOs.
\end{abstract}

\setboolean{displaycopyright}{false} 

\begin{document}
\nolinenumbers
\maketitle

\section{Introduction}
Continuous wave (CW) optical parametric oscillators (OPOs) have made significant progress since their demonstration \cite{1} and have been applied in various fields, such as atomic and molecular spectroscopy \cite{2}, terahertz generation \cite{3} and optical frequency division \cite{4}. One of its key advantages over lasers is its broader tuning range, enabling it to cover wavelengths that are difficult to reach with lasers \cite{5}. In particular, the doubly resonant OPO (DRO) has attracted much attention because of its lower threshold and narrower linewidths \cite{6}. The limiting factors for DRO tuning are energy conservation, phase matching, and the resonant cavity that provides gain \cite{7}. 
Research on OPO tuning properties encompasses not only the accessible wavelength range but also the tuning methods and their corresponding tuning rates. Different cavity configurations (such as a monolithic cavity\cite{7,8}, a semi-monolithic cavity\cite{9}, a two-mirror standing-wave cavity \cite{5}, and a four-mirror ring cavity \cite{10}), phase matching schemes (such as birefringent phase matching \cite{7,8,9} and quasi-phase matching \cite{5,10}), and nonlinear crystals (such as lithium niobate \cite{7,9} and potassium titanyl phosphate \cite{5,8,10}) provide distinct tuning approaches and lead to different tuning performances. 
Exploring the tuning properties under the aforementioned configurations has always been a crucial aspect of OPO research.

Backward OPOs (BOPOs) \cite{11} realized by backward quasi-phase matching (BQPM) exhibit unique tuning characteristics compared with traditional forward OPOs \cite{5,6,7,8,9,10}. Although BOPO output can be achieved without a resonant cavity due to its inherent feedback mechanism, the following problems exist simultaneously. A mirrorless OPO (MOPO) requires an extremely high threshold, which can usually only be achieved with pulsed pumping \cite{12,13,14,15,16}. At the high threshold, competition with stimulated Raman scattering \cite{12} and potential material damage can occur, especially in nonlinear materials where submicron period poling techniques have not yet been widely developed. The high threshold also limits the application of BOPOs in CW operations. 
In MOPOs operated in the pulsed regime, the verified tuning behavior is that changing the temperature or angle of the crystal has little effect on the output wavelength, which remains limited to approximately one hundred GHz \cite{12,13}. Varying the pump wavelength exerts only a small effect on the tuning of the narrowband backward wave. In contrast, energy conservation allows significant adjustment of the forward-wave wavelength. Moreover, the forward-wave spectrum inherits the characteristics of the pump spectrum, leading to a broad spectral width \cite{12,15}.
By analogy to a doubly resonant forward OPO (DRFO) \cite{17}, if a crystal satisfying BQPM is placed inside a cavity resonant with both the signal and idler waves—i.e., forming a doubly resonant BOPO (DRBO)—and pumped with a narrowband CW source, the threshold is expected to be reduced and a narrowband forward wave can be generated. Its wavelength can be significantly tuned by varying the pump wavelength, with fine adjustment achievable by changing the temperature.
Furthermore, owing to the narrow wavelength bandwidth of BQPM \cite{18,19}, the number of longitudinal modes within the gain bandwidth in a DRBO is expected to be significantly reduced compared with that in a DRFO \cite{7,18}.
Therefore, generating a DRBO with a narrowband CW pump is expected to constitute a novel type of OPO with unique characteristics and a practical solution to the aforementioned issues encountered under pulsed pumping.
However, current experimental implementations of BOPO are limited to the mirrorless pulsed pumping scenario, with CW pumping remaining entirely theoretical \cite{20}. 
Clearly, experimental verification of the generation and characteristics of a DRBO under continuous-wave pumping—particularly its tuning behavior—is of critical importance. This would not only perfect the implementation of BQPM in OPOs but also expand the tuning capabilities of CW OPOs and broaden their potential applications, such as serving as a strong candidate for atmospheric greenhouse gas detection in the mid-infrared region \cite{20}.

Here, we report the realization of a DRBO with CW pumping using BQPM. A semi-monolithic cavity configuration was employed, utilizing a periodically poled potassium
titanyl phosphate (PPKTP) crystal satisfying 7th-order Type-0 BQPM \cite{19}, with the generated signal and idler waves resonating within the cavity. The tuning properties were characterized by scanning the cavity length. The pump tuning range was approximately 13.66 nm, resulting in tuning ranges of 56.85 nm and 0.89 nm for the forward and backward waves, with corresponding frequency tuning rates of 1.02 and -0.02, respectively. The crystal temperature was adjusted over a range of 20°C, resulting in tuning ranges of 59.7 GHz and 59.4 GHz for the forward and backward waves, with corresponding temperature tuning rates of 3 GHz/K and -3 GHz/K, respectively.

\section{THEORY MODEL}
First, the oscillation threshold of a DRBO is given, and the two resonant wavelengths are non-degenerate.

The coupled-amplitude equations for backward phase matching can be written as \cite{21,22,23}:

\begin{equation}
\begin{array}{l}
\frac{{d{A_1}}}{{dz}} =  - \frac{{2i{d_{eff}}\omega _1^2}}{{{k_1}{c^2}}}{A_3}A_2^*{e^{i\Delta kz}}\\
\frac{{d{A_2}}}{{dz}} = \frac{{2i{d_{eff}}\omega _2^2}}{{{k_2}{c^2}}}{A_3}A_1^*{e^{i\Delta kz}}\\
\frac{{d{A_3}}}{{dz}} = \frac{{2i{d_{eff}}\omega _3^2}}{{{k_3}{c^2}}}{A_1}{A_2}{e^{ - i\Delta kz}}
\end{array},\label{eq1}
\end{equation}
where $\Delta k = {k_3} - {k_2} + {k_1}$ is the phase-matching condition, ${A_q}$ is the amplitude of the wave $q$, with $q = 1,{\rm{ }}2,{\rm{ }}3$ corresponding to the backward, forward, and pump waves, respectively; ${\omega _q}$ and ${k_q}$ are respectively the angular frequency, and wave vector of the wave $q$; ${d_{eff}}$ is the effective nonlinear coefficient; $c$ is the light velocity in vacuum.

Next, consider the case where the pump wave ${A_3}$ is constant and the phase matching is satisfied $\Delta k = 0$. By solving \eqref{eq1}, we obtain:
\begin{equation}
\begin{array}{l}
{A_1}(z) = \frac{{{A_1}(L)}}{{\cos (gL)}}\cos (gz) + \frac{{i\sqrt {\frac{{{n_2}{\omega _1}}}{{{n_1}{\omega _2}}}} \frac{{{A_3}}}{{\left| {{A_3}} \right|}}A_2^*(0)}}{{\cos (gL)}}\sin [g(L - z)]\\
{A_2}(z) = \frac{{{A_2}(0)}}{{\cos (gL)}}\cos [g(L - z)] + \frac{{i\sqrt {\frac{{{n_1}{\omega _2}}}{{{n_2}{\omega _1}}}} \frac{{{A_3}}}{{\left| {{A_3}} \right|}}A_1^*(L)}}{{\cos (gL)}}\sin (gz)
\end{array},\label{eq2}
\end{equation}
where ${n_j}$ is the refractive index of the wave $j$; $L$ is the length of the nonlinear crystal; $g$ satisfies the equation:
\begin{equation}
{g^2} = \frac{{4d_{eff}^2\omega _1^2\omega _2^2}}{{{k_1}{k_2}{c^4}}}{\left| {{A_3}} \right|^2}.\label{eq3}
\end{equation}

Following the assumptions of reference \cite{21}, the nonlinear crystal itself acts as a cavity whose two ends (denoted as the '$in$' and '$out$' ends) have reflection coefficients ${r_{1j}},{\rm{ }}{r_{2j}}$, $j = in,{\rm{ }}out$ to the backward wave ${A_1}$ and forward wave ${A_2}$, respectively, with the corresponding mirror reflectivities being ${R_{1j,2j}} = {\left| {{r_{1j,2j}}} \right|^2}$. Both ends remain transparent to the pump wave ${A_3}$. Using \eqref{eq2} and the derivation approach from \cite{21} (the self-consistent condition requires that the fields ${A_1}$ and $A_2^*$ reproduce themselves after they complete round trip inside the cavity.), we can obtain the oscillation condition of a DRBO:
\begin{equation}
\begin{array}{l}
\left[ {{r_{1in}}{r_{1out}}\frac{{{e^{i2{k_1}L}}}}{{\cos (gL)}} - 1} \right]\left[ {r_{2in}^*r_{2out}^*\frac{{{e^{ - i2{k_2}L}}}}{{\cos (gL)}} - 1} \right]\\
{\rm{                                                               }} = {r_{1in}}{r_{1out}}r_{2in}^*r_{2out}^*{e^{ - i2({k_2} - {k_1})L}}{\tan ^2}(gL)
\end{array}.\label{eq4}
\end{equation}
The minimum threshold gain \cite{21} will result the relations:
\begin{equation}
\left\{ {\begin{array}{*{20}{c}}
{2{k_1}L + \frac{{{\varphi _{1in}}}}{2} + \frac{{{\varphi _{1out}}}}{2} = 2m\pi }\\
{2{k_2}L + \frac{{{\varphi _{2in}}}}{2} + \frac{{{\varphi _{2out}}}}{2} = 2n\pi }
\end{array}} \right.,\label{eq5}
\end{equation}
where $m,{\rm{ }}n$ are two integers and
\begin{equation}
\left\{ {\begin{array}{*{20}{c}}
{r_{1j}^2 = {R_{1j}}{e^{i{\varphi _{1j}}}}}\\
{{{(r_{2j}^*)}^2} = {R_{2j}}{e^{ - i{\varphi _{2j}}}}}
\end{array}} \right..\label{eq6}
\end{equation}
Using \eqref{eq5} and \eqref{eq6} in \eqref{eq4}, we obtain the result
\begin{equation}
\cos (gL) = \frac{{\sqrt {{R_{1in}}{R_{1out}}}  + \sqrt {{R_{2in}}{R_{2out}}} }}{{1 + \sqrt {{R_{1in}}{R_{1out}}{R_{2in}}{R_{2out}}} }}. \label{eq7}
\end{equation}
\eqref{eq7} can be rewritten as
\begin{equation}
g = \frac{{\arccos (R)}}{L},{\rm{ }}R = \frac{{\sqrt {{R_{1in}}{R_{1out}}}  + \sqrt {{R_{2in}}{R_{2out}}} }}{{1 + \sqrt {{R_{1in}}{R_{1out}}{R_{2in}}{R_{2out}}} }}.\label{eq8}
\end{equation}
By combining \eqref{eq3}, \eqref{eq8}, and ${I_3} = 2{n_3}{\varepsilon _0}c{\left| {{A_3}} \right|^2}$(${I_3}$ is the intensity of the pump wave and ${\varepsilon _0}$ is the vacuum dielectric constant), we find that the threshold intensity is given by
\begin{equation}
{I_{th}} = {I_3} = \frac{{{n_1}{n_2}{n_3}c{{[\arccos (R)]}^2}}}{{2d_{eff}^2{L^2}{\mu _0}{\omega _1}{\omega _2}}}.\label{eq9}
\end{equation}

Next, the relationship between the DRBO threshold and that of a MOPO is examined. By setting ${R_{1j}} = {R_{2j}} = 0$ (i.e., $R = 0$) in \eqref{eq9}, the expression reduces to the threshold of a MOPO \cite{22}. Therefore, the introduction of a cavity reduces the threshold required to realize a BOPO. In the case of a DRFO, however, setting the mirror reflectivity to zero invalidates the approximation involving the cosh function \cite{21}, which results in an infinite threshold. In a similar manner, the threshold for a singly resonant BOPO can be derived.

Then the bandwidth of backward phase matching is derived. Using energy conservation, the phase-matching condition, and $\Delta kL = \pm 0.886\pi$, the bandwidth is obtained as \cite{18}
$\Delta {\omega _b} = {{1.772\pi } \mathord{\left/
 {\vphantom {{1.772\pi } {(v_1^{ - 1} + v_2^{ - 1})}}} \right.
 \kern-\nulldelimiterspace} {(v_1^{ - 1} + v_2^{ - 1})}}$,
where 
\begin{math}
{v_q} = {{\partial {\omega _q}} \mathord{\left/
 {\vphantom {{\partial {\omega _q}} {\partial {k_q}}}} \right.
 \kern-\nulldelimiterspace} {\partial {k_q}}},
\end{math}
and ${\omega _3}$ is treated as a constant in this process. Compared with the forward case, the compression factor of the bandwidth is given by $(v_1^{-1} + v_2^{-1}) / \left| v_1^{-1} - v_2^{-1} \right|$ \cite{18}.
 
Finally, the expressions for the wavelength and temperature tuning rates are given. The wavelength tuning rate is \cite{12}, 
\begin{equation}
\frac{{\partial {\omega _1}}}{{\partial {\omega _3}}} = \frac{{v_2^{ - 1} - v_3^{ - 1}}}{{v_1^{ - 1} + v_2^{ - 1}}},  \frac{{\partial {\omega _2}}}{{\partial {\omega _3}}} = \frac{{v_1^{ - 1} + v_3^{ - 1}}}{{v_1^{ - 1} + v_2^{ - 1}}},\label{eq10}
\end{equation}
and the temperature tuning rate is \cite{24}, 
\begin{equation}
\frac{{\partial {\nu _1}}}{{\partial T}} =  - \frac{{\partial {\nu _2}}}{{\partial T}} = \frac{1}{{{n_1} + {n_2}}}[{\nu _2}(\frac{{\partial {n_2}}}{{\partial T}} - \frac{{\partial {n_3}}}{{\partial T}}) - {\nu _1}(\frac{{\partial {n_1}}}{{\partial T}} + \frac{{\partial {n_3}}}{{\partial T}})],\label{eq11}
\end{equation}
where ${\nu _q}$ denotes the frequency of the wave $q$ and the thermal expansion coefficient \cite{24} is neglected.

\section{EXPERIMENTAL RESULTS}
\begin{figure*}[ht]
\centering\includegraphics[width=17cm]{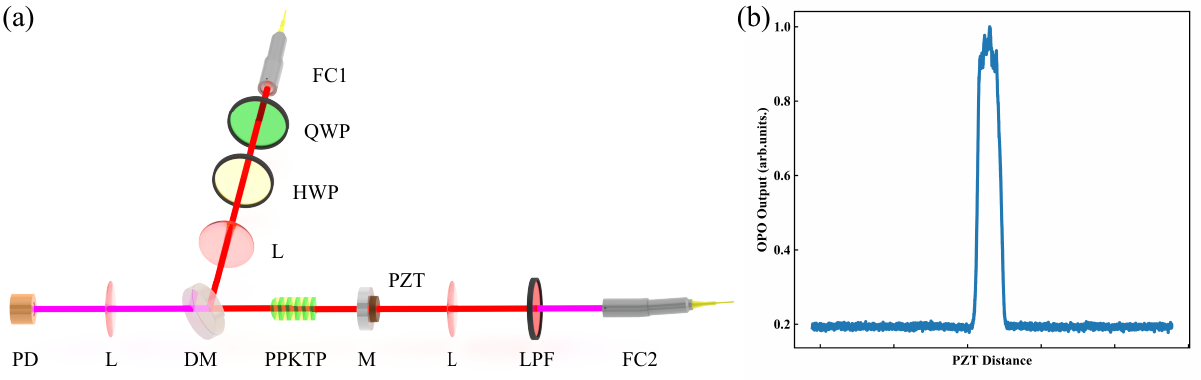}
\caption{(a) Experimental setup. FC: fiber coupler; Q(H)WP: quarter- (half-) wave plate; L: lens; DM: dichroic mirror;  PPKTP: periodically poled KTP crystal; M: concave mirror; PZT: piezoelectric transducer; LPF: long-pass filter; PD: photodetector. (b) Oscilloscope trace of an OPO output obtained during cavity-length scanning.}
\label{fig1}
\end{figure*}

The experimental setup is shown in Fig.\ref{fig1}(a). The pump laser is a CW from a Ti: Sapphire laser. It is coupled into a single-mode fiber and outputs through FC1 to pump the OPO. QWP and HWP are used to adjust the pump polarization to vertical. The pump laser is mode matched to the OPO through a lens. 
The OPO adopts a semi-monolithic cavity configuration, comprising a PPKTP crystal with a convex surface of 8 mm radius of curvature on one end and a flat surface on the other, together with an external concave mirror of 15 mm radius of curvature.
The convex surface of the crystal serving as the input coupler has a coating with an antireflection at 780-795 nm, a high reflection at 1550-1630 nm (\textit{R}>99.9\%) and the flat surface has a coating with an antireflection at both wavelengths. The concave mirror serving as the output coupler has a coating with a partial reflection (\textit{R}$ \approx $99.2\%) at 1550-1630 nm and an antireflection at 780-795 nm. The dimensions of the crystal are $\rm{1 mm \times2 mm \times9 mm}$, and its temperature was controlled. The poling period of the crystal is 2.95 µm and the 7th order Type-0 BQPM is achieved. A piezoelectric transducer (PZT) is mounted on the concave mirror to scan the cavity length. The OPO is a single-pass configuration for the pump. When the threshold is reached, it starts to oscillate and generates the forward propagating idler and backward propagating signal. Owing to the limitation of the standing-wave cavity, both the signal and idler are output from the output coupler.
After being collimated by a lens, the signal and idler beams are coupled into FC2. Before entering FC2, the pump beam is removed by a LPF. FC2 is connected to an optical spectrum analyzer (OSA, not shown in the figure).
Because the reflectivity of the crystal’s convex surface is less than 100\%, a small fraction of the signal and idler leaks from this surface. After passing through a DM and a lens, the leaked beams are incident on a PD connected to an oscilloscope (not shown). An example of the OPO output observed on the oscilloscope during cavity-length scanning is shown in Fig.\ref{fig1}(b).

\begin{figure}[htb]
\centering\includegraphics[width=8.5cm]{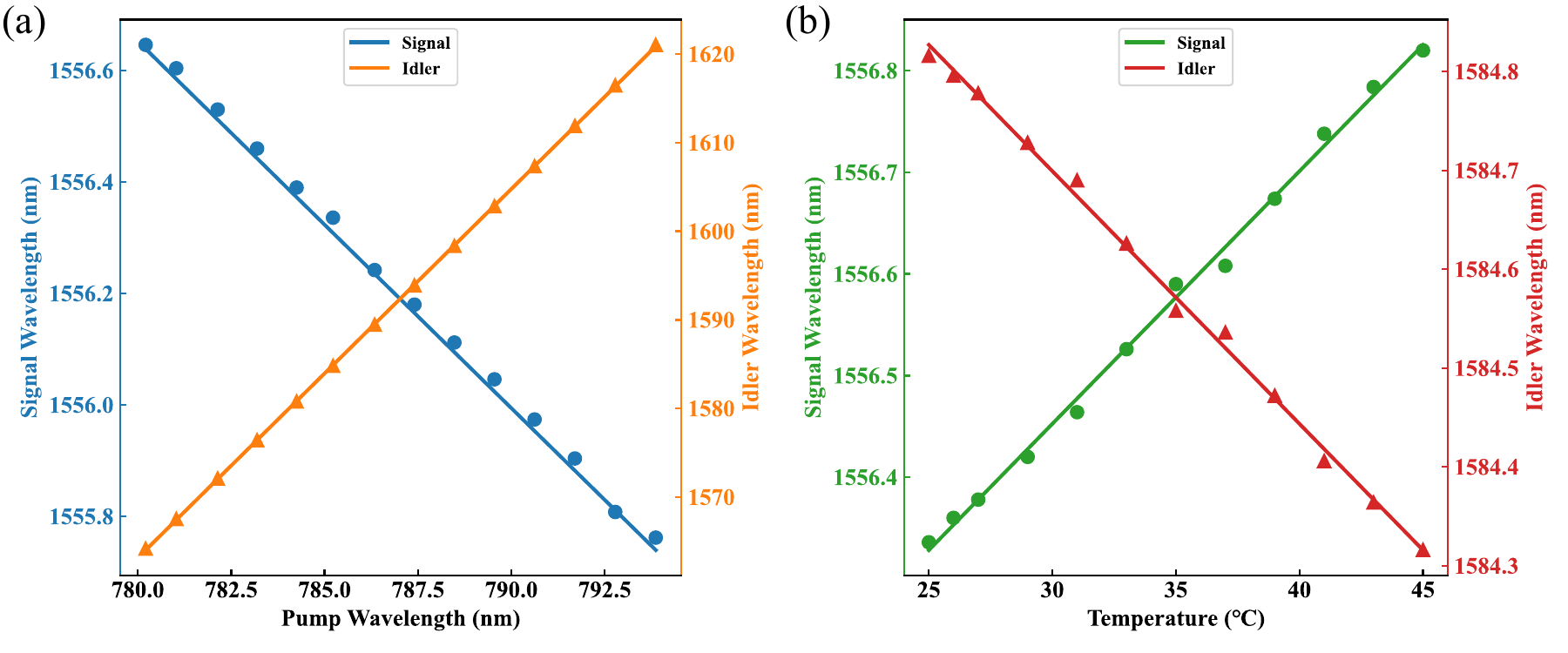}
\caption{Dependence of the signal and idler wavelengths on (a) the pump wavelength and (b) the crystal temperature. Circles and triangles denote the experimental data for the backward-propagating signal and forward-propagating idler waves, respectively, while the solid lines represent the corresponding linear fitting results.}
\label{fig2}
\end{figure}

First, the wavelength of the pump laser was set to 785.23 nm, and the crystal temperature was maintained at 25°C. 
We provide a theoretical estimate of the threshold under these conditions. Assuming that the threshold intensity and threshold power satisfy \cite{21,25} ${I_{th}} = {{{P_{th}}} \mathord{\left/
 {\vphantom {{{P_{th}}} {\pi {w^2}}}} \right.
 \kern-\nulldelimiterspace} {\pi {w^2}}}$, where $w$ is the spot radius.
 By assuming a beam radius of $w = 35$ µm and a nonlinear coefficient of ${d_{33}} = 4.8$ $\rm pm/V$ \cite{26}, and using the refractive index data from Ref.\cite{27} together with the other experimental parameters, substitution into \eqref{eq9} gives the threshold power of approximately 2.6 W.
In the experiment, the cavity length was scanned using the PZT while the pump power was gradually increased to approximately 3.1 W. At this point, the OPO output idler and signal waves were observed on the OSA, with wavelengths of 1584.82 nm and 1556.34 nm, respectively. The corresponding wavelengths obtained from numerical simulation \cite{19} are 1584.84 nm and 1556.34 nm, showing excellent agreement with the experimental results.

Next, we characterized the dependence of the signal and idler wavelengths on the pump wavelength. The crystal temperature was set to 25°C, the pump power was increased to approximately 3.4 W, and the cavity length scan was maintained. The pump wavelength was tuned from 780.21 nm to 793.87 nm (a total change of 13.66 nm). The idler wavelength detected on the OSA shifted from 1564.19 nm to 1621.04 nm (a total change of 56.85 nm), while the signal wavelength shifted from 1556.65 nm to 1555.76 nm (a total change of 0.89 nm). The specific experimental results are shown in Fig.\ref{fig2} (a). After linear fitting of the experimental data, the tuning rates of the idler and signal waves were determined as follows:
\begin{math}
\partial \nu_2 / \partial \nu_3 =1.02, \, \partial \nu_1 / \partial \nu_3 =-0.02,
\end{math}
 respectively. 
Using \eqref{eq10}, when the pump wavelength is 785.23 nm, the calculated wavelength tuning rates of the idler and signal waves are approximately 1.02 and –0.02, respectively, in good agreement with the experimental data.
It can be seen that varying the pump wavelength leads to a significant shift in the forward-wave wavelength, while the backward-wave wavelength exhibits only a small variation.

Then we characterized its temperature tuning properties. The pump power was kept constant at approximately 3.4 W, the pump wavelength was set to 785.23 nm, and the cavity length scan remained unchanged. The crystal temperature was varied from 25°C to 45°C, resulting in a corresponding change in the idler wavelength from 1584.82 nm to 1584.32 nm ( a total change of 59.7 GHz), and a shift in the signal wavelength from 1556.34 nm to 1556.82 nm (a total change of 59.4 GHz). The results are shown in Fig.\ref{fig2} (b). The linear fitting curves reveal that the temperature tuning rates of the signal and idler waves are related as follows: 
\begin{math}
\partial \nu_2 / \partial T =- \partial \nu_1 / \partial T =3\, {\rm GHz/K}.
\end{math}
Using \eqref{eq11} together with refractive index data from Ref.\cite{28}, at 25°C the calculated temperature tuning rates of the idler and signal waves are approximately 3.9 GHz/K and –3.9 GHz/K, respectively, also in good agreement with the experimental data.
It is also evident that temperature variation produces nearly identical and relatively small effects on both the forward and backward waves. The tuning behavior of the DRBO observed experimentally under CW pumping is similar to that of the MOPO under pulsed pumping \cite{12}, as both are governed by BQPM constraints.

\begin{figure}[htb]
\centering\includegraphics[width=8.5cm]{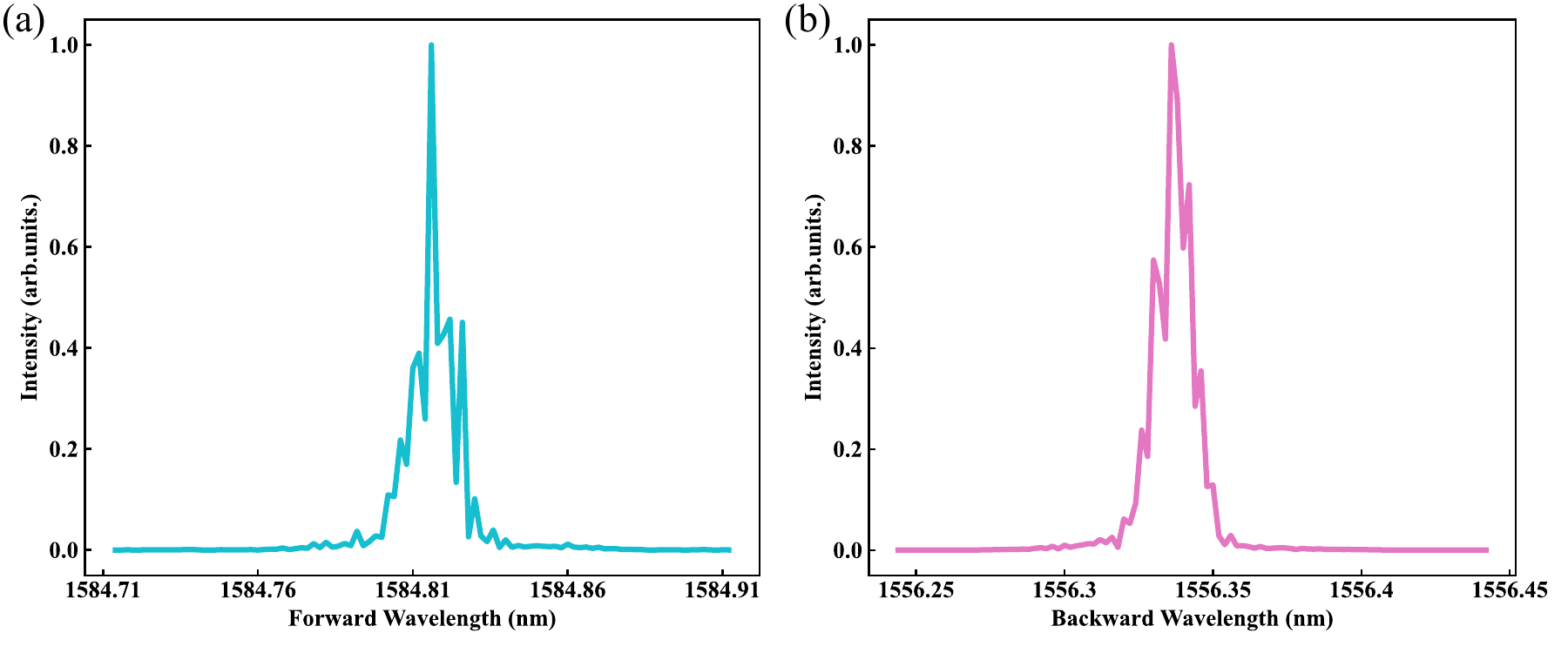}
\caption{Spectra of the (a) forward and (b) backward waves measured with the optical spectrum analyzer.}
\label{fig3}
\end{figure}

Finally, the spectra of the forward and backward waves observed with the OSA during cavity-length scanning are presented in Fig.\ref{fig3}. Both spectra are limited by the OSA resolution of 0.02 nm. These features differ from those of MOPOs under pulsed pumping \cite{12}; in contrast, under CW pumping both the backward and forward waves exhibit narrowband.

\section{Conclusion}
In conclusion, we experimentally employed a semi-monolithic cavity to investigate the tuning properties of the DRBO under CW pumping by monitoring the output during PZT-based cavity scanning. These measurements clearly reveal the distinctive tuning behavior in comparison with the DRFO.
By adjusting the CW pump wavelength, we were able to generate a backward wave with a nearly constant wavelength and a forward wave with a significantly varying wavelength. Both waves exhibited low sensitivity to temperature tuning. For the forward wave, varying the pump wavelength and temperature enabled coarse and fine tuning of the OPO output, respectively. The backward wave, on the other hand, is well-suited for applications that require a stable output, insensitive to variations in the pump wavelength and temperature, or for scenarios that demand high-precision tuning. 
Although active cavity stabilization was not implemented in the present work, the demonstrated results provide unambiguous evidence of the oscillation process and the underlying tuning mechanism. Achieving long-term stabilized operation is expected to be feasible with optimized cavity design, the use of crystals supporting lower-order BQPM (and hence a reduced threshold), and the implementation of active locking techniques, which will be the subject of future work.

\begin{backmatter}
\bmsection{Funding}
 National Key Research and Development Program of China (2022YFB3903102, 2022YFB3607700), National Natural Science Foundation of China (NSFC)(62435018), Innovation Program for Quantum Science and Technology (2021ZD0301100), USTC Research Funds of the Double First-Class Initiative(YD2030002023), and Research Cooperation Fund of SAST, CASC (SAST2022-075).

\bmsection{Disclosures}
The authors declare no conflicts of interest.

\bmsection{Data availability}
Data underlying the results presented in this paper are not publicly available at this time but may be obtained from the authors upon reasonable request.

\end{backmatter}

\bibliography{References}

\end{document}